# Tuning the magnetic properties of $Fe_{50-x}Mn_xPt_{50}$ thin films

*Ezhil A. Manoharan,[1] Gary Mankey[1] and Yang-Ki Hong[2]*

[1]*Department of Physics, University of Alabama and MINT, Tuscaloosa, 35401, USA*

[2]*Department of Electrical and Computer Engineering and MINT, University of Alabama, Tuscaloosa, 35401, USA*

**Abstract**

The magnetic and structural properties of highly ordered (S ~ 0.82) epitaxial $Fe_{50-x}Mn_xPt_{50}$ thin films were investigated. We report the change in the magnetic properties of Mn doped FePt epitaxial thin films. This study differs from the earlier experimental studies on Mn doped FePt based alloys. Ordered $L1_0$ $Fe_{50-x}Mn_xPt_{50}$ (x=0, 6, 9, 12 and 15) thin films with a constant thickness of 45 nm were prepared by co-sputtering $Fe_{50}Pt_{50}$ and $Mn_{50}Pt_{50}$ on to MgO (100) single crystal substrate. We find a significant increase in the coercivity for Fe-Mn-Pt thin films. We have shown that this increase in magnetic properties coincide with the tetragonal distortion, while the recent first principles study of Mn doped FePt showed the sub lattice ordering of ferromagnetically aligned Mn atoms would lead to increase in magnetic properties in the FeMnPt ternary alloy system with fixed Pt concentration. At x=12 the coercivity has increased by 46.4 % when compared to $Fe_{50}Pt_{50}$. The increase in magnetic properties in $Fe_{50-x}Mn_xPt_{50}$ is due to the tetragonal distortion as experimental c/a ratio is larger than the expected c/a ratio for ferromagnetically ordered Mn atoms in the sublattice at the concentration x=12. Thus we show that high temperature deposition and high temperature annealing is one of the methods to achieve large coercivity in Mn doped FePt as it leads to tetragonal distortion.

# I. Introduction

Ordered L1$_0$ FePt has high magnetocrystalline anisotropy energy [1-3] and hence it's a potentially promising candidate for future recording technologies like heat assisted magnetic recording (HAMR) [4]. The L1$_0$ unit cell is a tetragonal distorted face centered cubic cell with the c- axis lie perpendicular to the plane of the base. The c/a ratio is one of the most important structural parameter to tune the magnetic properties of the ferromagnetic thin films as it leads to changes in electronic arrangements which may alter the magnetic properties. In L1$_0$ FePt, the high anisotropy is due to the combination of tetragonal distortion, exchange effects and spin orbit moment of Pt [5]. It was predicted that maximum anisotropy for FePt would be achieved at c/a ratio close to experimental value of 0.96 [6]. Structural changes can be a very important in tuning the magnetic properties of ferromagnetic films like coercivity and magnetization. Ferromagnetic properties like coercivity were shown to change with respect to the thickness of the thin film in Fe$_{50}$Pt$_{50}$ [7-11]. It was shown that by exploiting the tetragonal distortion, magnetic properties like magneto crystalline anisotropy can be maximized [12]. High temperature deposition and high temperature annealing with longer annealing time were known to induce structural changes ( tetragonal phase ) in ordered L1$_0$ Fe$_{50}$Pt$_{50}$ [13-16]. These strucutural changes contributed to the increase in the coercivity in the L1$_0$ Fe$_{50}$ Pt$_{50}$ [17]. Thus tetragonal distortion is one of the important factors in maximizing the magnetic properties like magneto crystalline anisotropy and curie temperature, both are very important factors for HAMR, as it temporarily lowers the coercivity by heating the local area of the recording media above the Curie temperature. It was suggested that to achieve 4 Tb/ in$^2$, a saturation magnetization of 800 emu/ cc, magnetic crystalline anisotropy of 5 x 10$^6$ J/m$^3$ and a curie temperature of 700-750 K is required [18]. L1$_0$ FePt is doped with non magnetic species such as Ni, Mn, or Cu permits the tuning of magnetic properties of tertary FePt system as the composition of Fe decreases and the doped element increase in composition while keeping the Pt composition constant [19-24]. L1$_0$ Fe$_{50-x}$Mn$_x$Pt$_{50}$ is a magnetic chameleon system since with small change in the Mn concentration could lead to a significant change in the magnetic properties [25], which make them ideal candidates for thin film recording media, and developing methods to control the magnetization, Curie temperature and anisotropy are desired to engineer the material for optimal properties for the future technologies [26]. A calculation of the magnetic phase diagram for Fe$_{50-x}$Mn$_x$Pt$_{50}$ published in 2015 [25], as the concentration of Mn changes with fixed Pt concentration, the magnetic phases change from collinear ferromagnetic phase to non collinear ferromagnetic phase to non collinear antiferromagnetic phase to collinear antiferromagnetic phase. Fe rich Fe$_{50-x}$Mn$_x$Pt$_{50}$ alloys are promising materials for heat assisted magnetic recording, and thus understanding the temperature dependence of these alloys is very important to predict their thermomagnetic properties with the focus on future technologies. The ideal candidates for HAMR will have a high anisotropy and a relatively low Curie temperature. Burkert et al [26] predicted that for Fe$_{50-x}$Mn$_x$Pt$_{50}$ system the magnetic crystalline anisotropy energy increases up to 33 % for the concentration x = 12.5 and the effect was attributed to band filling. This was a significant prediction as the increase in anisotropy could be very important for the future magnetic recording applications but the previous experimental results for Fe$_{50-x}$Mn$_x$Pt$_{50}$ on MgO showed significant decrease in magnetic properties with the increase in the Mn concentration [28-31]. Fe$_{50-x}$Mn$_x$Pt$_{50}$ thin films on a-plane Al$_2$O$_3$ also showed a significant decrease in magnetic properties [32]. Meyer [29] attributed the significant reduction of magnetization and anisotropy due to the antiparallel alignment of Mn moments as observed in circular x-ray magnetic dichroism. In 2016, Cuadrado et al. [33] explained that in L1$_0$ Fe$_{50-x}$Mn$_x$Pt$_{50}$ bulk alloy with fixed Pt concentration, Mn atoms align in two different magnetic sublattice ordering, one is ferromagnetic alignment (FM) and the other is antiferromagnetic alignment (AFM). The authors demonstrated that FM alignment will enhance and AFM alignment will reduce the magnetic anisotropy. Thus previous experimental results in $Fe_{50-x}Mn_xPt_{50}$ with reduction in magnetic properties were attributed to the AFM alignment of the Mn atoms. In this work we showed that it is possible to increase the magnetic properties like coercivity by tetragonal distortion due to high temperature deposition and high temperature annealing with longer annealing time.

## II. Experiment

Epitaxial $Fe_{50-x}Mn_xPt_{50}$ (x=0, 6,9,12 and 15) thin films were prepared by co-sputtering $Fe_{50}Pt_{50}$ and $Mn_{50}Pt_{50}$ on to MgO (100) single crystal substrate at 3.5 mTorr Ar and the thickness was fixed at 45 nm. The thin films were directly deposited on the MgO substrate at 780°C to avoid diffusion and the samples were annealed at 920°C for 1 hr. Before deposition the substrates were sputter cleaned at 0.5 mTorr Ar for 5 mins. The films are produced in a UHV sputtering system with RHEED and Auger electron spectroscopy with the starting pressure better than 6 x $10^{-10}$ torr. RHEED is used to verify epitaxy. Figure 1 shows the 220 pole figure for the different $Fe_{50-x}Mn_xPt_{50}$ thin films where the four-fold symmetry can be seen. XRD 2θ- θ scans for different concentrations (x=0, 6,9,12 and 15) are shown in the figure 2. The shift in the 001 and 002 peaks shows the expansion of c-axis. For x=0 and x=6, the XRD scans shows singlet 002 peaks, while for x= 9, 12 and 15 the XRD scans shows the splitting of 200 peaks due to tetragonal distortions, as the reflection from peaks like 002 and 200 are no longer equivalent in the tetragonal structure. The distance d between the planes (h,k,l) in a tetragonal lattice is given by

$$\frac{1}{d^2} = \frac{h^2+k^2}{a^2} + \frac{l^2}{c^2}$$

Substituting this in the Bragg's equation, we get

$$2d \cdot \sin\theta = n\lambda$$

$$4d^2 \sin^2\theta = n^2\lambda^2$$

$$\frac{4 \cdot \sin^2\theta}{\lambda^2} = \frac{h^2+k^2}{a^2} + \frac{l^2}{c^2}$$

The out of plane $c$ and in plane lattice parameter $a$ were solved using the doublet peaks, while for the singlet peaks out of plane $c$ were solved from 002/001 peaks and the out of plane lattice parameter was used to calculate the in plane lattice parameter $a$ in the tetragonal lattice for 202 peaks, $c$ and $a$ values are summarized in the Table I. The intense 001 peaks suggest all the samples are characteristic of the chemically ordered $L1_0$ structure. The chemical ordering S for these thin films are found to between 0.82 to 0.84. RHEED was used to check the epitaxial nature of the thin films during the deposition, Figure 3 shows the RHEED patterns for different concentrations (x=0, 6,9,12 and 15). The RHEED pattern agrees with the XRD results, as the RHEED pattern is different for the thin films samples with tetragonal distortion, for the concentrations x=0 and x= 6, the RHEED patterns with no tetragonal distortion are different from the concentration x=9, 12 and 15 with tetragonal distortion.

## III. Results

Figure 4 shows the out of plane hysteresis measurement at 300 K. For the two films x=0 and x= 6, the magnetization is consistent with the previous results but x=9 shows increase in the coercivity which coincides with the onset of the 200 peak. The films x= 12 and 15 shows significant increase in both magnetization and coercivity which is different from the previous results, these films show prominent peak splitting due to tetragonal distortion. The decrease in the magnetic properties due to addition of Mn ( x= 6) in FePt system could be attributed to the presence of AFM phase in Mn atoms as c/a ratio for this sample agree with the previous results, while the increase in the magnetic properties due to the addition of Mn could be attributed to the tetragonal distortion as per the XRD scans but not due to presence of FM phase, since the c/a ratio for these three samples are different from the predicted c/a ratio of the FM phase which is $\approx 0.971$. In the FM phase the c/a ratio decreases from 0.96 (x=0) to 0.94 (x=15). Figure 5 shows the out of plane saturation magnetization and coercivity as a function of composition where it can be see that the magnetic properties like magnetization and coercivity increase due to the onset of tetragonal distortion and attains maximum for the composition at x=12. For the concentration x=9, 12 and 15, there is about 39%, 46% and 13% increase in coercivity as compared to the FePt (x=0). The lattice parameters and magnetic properties are summarized in the table I.

## IV. Conclusion

We have shown that it possible to increase the magnetic properties of $Fe_{50-x}Mn_xPt_{50}$ with respect to FePt by inducing tetragonal distortion with high annealing temperature and high deposition temperature. This is one of the ways to increase the magnetic properties of $Fe_{50-x}Mn_xPt_{50}$ system apart from the ferromagnetic alignment of Mn atoms in the sublattice as proposed by Cuadrado et al. The original prediction for increased in magnetic properties in $Fe_{50-x}Mn_xPt_{50}$ system at x=12.5 coincides with our results. In order to understand the relevance of its use in HAMR, temperature dependence magnetic measurements at this concentration are needed, since HAMR is greatly affected by the Curie temperature, the thermomagnetic properties at the concentration of x= 12 will reveal more information about its curie temperature in the context of technologies like HAMR. $Fe_{38}Mn_{12}Pt_{50}$ may be a promising candidate for the next generation data storage media, since at x= 12, there is a significant increase in coercivity, this concentration may find potential applications in the data storage due to its increased magnetic anisotropy.

ACKNOWLEGEMENTS:

We acknowledge MINT support through shared facilities. This work was supported in part by the NSF-CMMI under award numbers 1463301.

TABLE I. Composition, out of plane and in plane lattice parameters, c/a ratio, saturation magnetization and coercivity of $Fe_{50-x}Mn_xPt_{50}$ films

| $Fe_{50-x}Mn_xPt_{50}$ | c | a | c/a | $M_s$ (300K) | $H_c$ |
|---|---|---|---|---|---|
| 0 | 3.711 | 3.875 | 0.958 | 1046 | 1890 |
| 6 | 3.715 | 3.865 | 0.961 | 1030 | 1753 |
| 9 | 3.719 | 3.842 | 0.968 | 990 | 2629 |
| 12 | 3.742 | 3.851 | 0.971 | 1140 | 2767 |
| 15 | 3.747 | 3.854 | 0.972 | 1061 | 2137 |

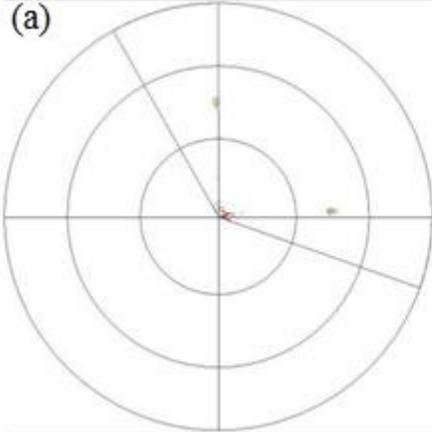
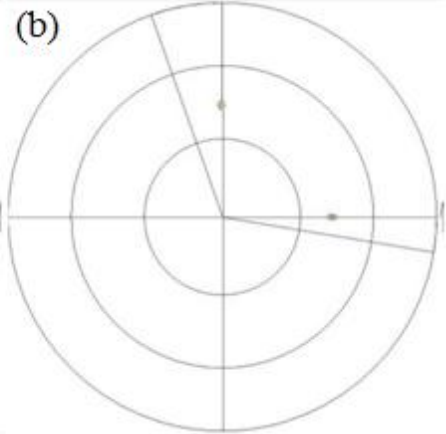
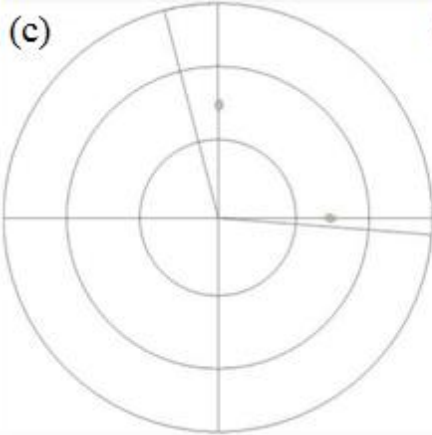
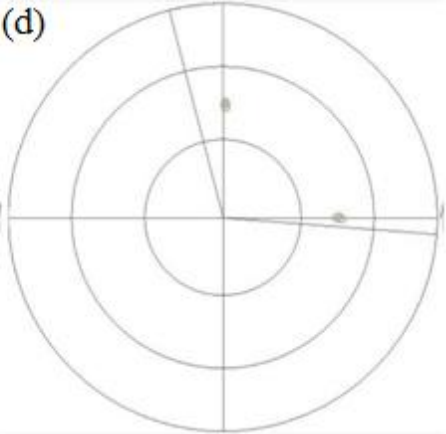
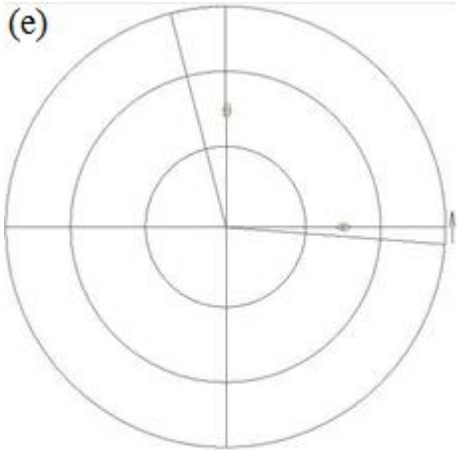

FIG. 1. 220 Pole figure for (a) x=0, (b) x=6, (c) x=9, (d) x= 12 and (e) x= 15 $Fe_{50-x}Mn_xPt_{50}$ thin films.

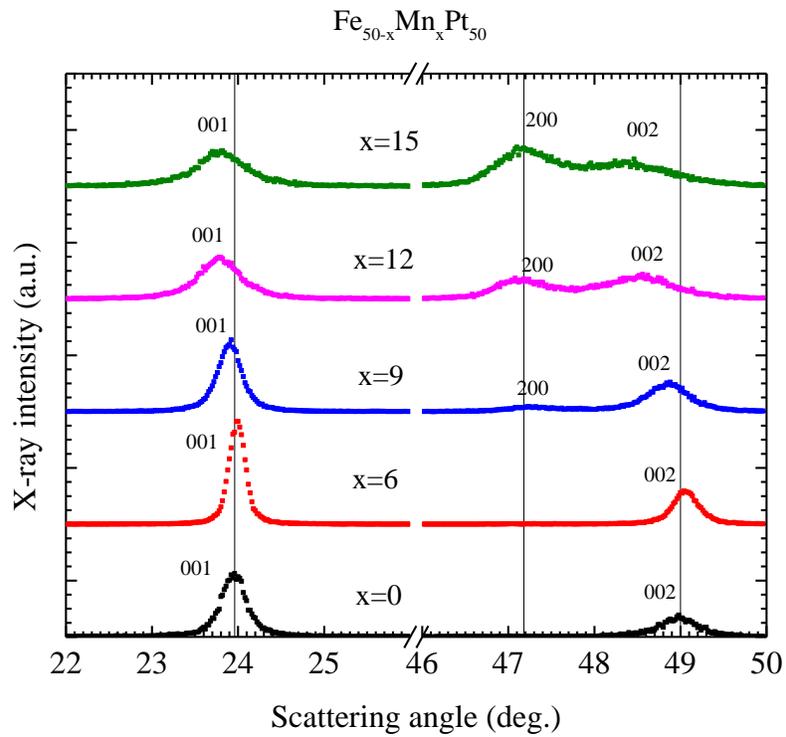

FIG. 2. XRD 2θ-θ scans of 45nm $Fe_{50-x}Mn_xPt_{50}$ thin films

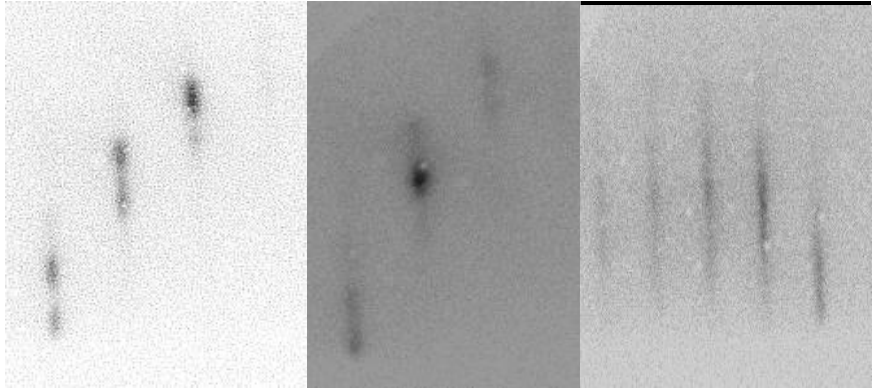

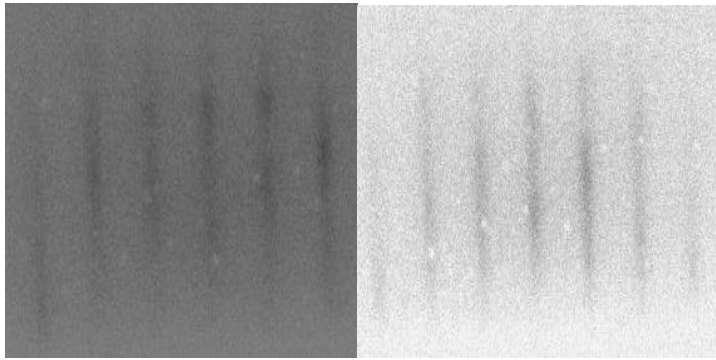

FIG.3. RHEED Pattern for $Fe_{50-x}Mn_xPt_{50,}$ top half ( from left to right ) i) x=0, ii) x= 6, iii) x= 9. Bottom half ( from left to right ) iv) x=12 and v) x= 15.

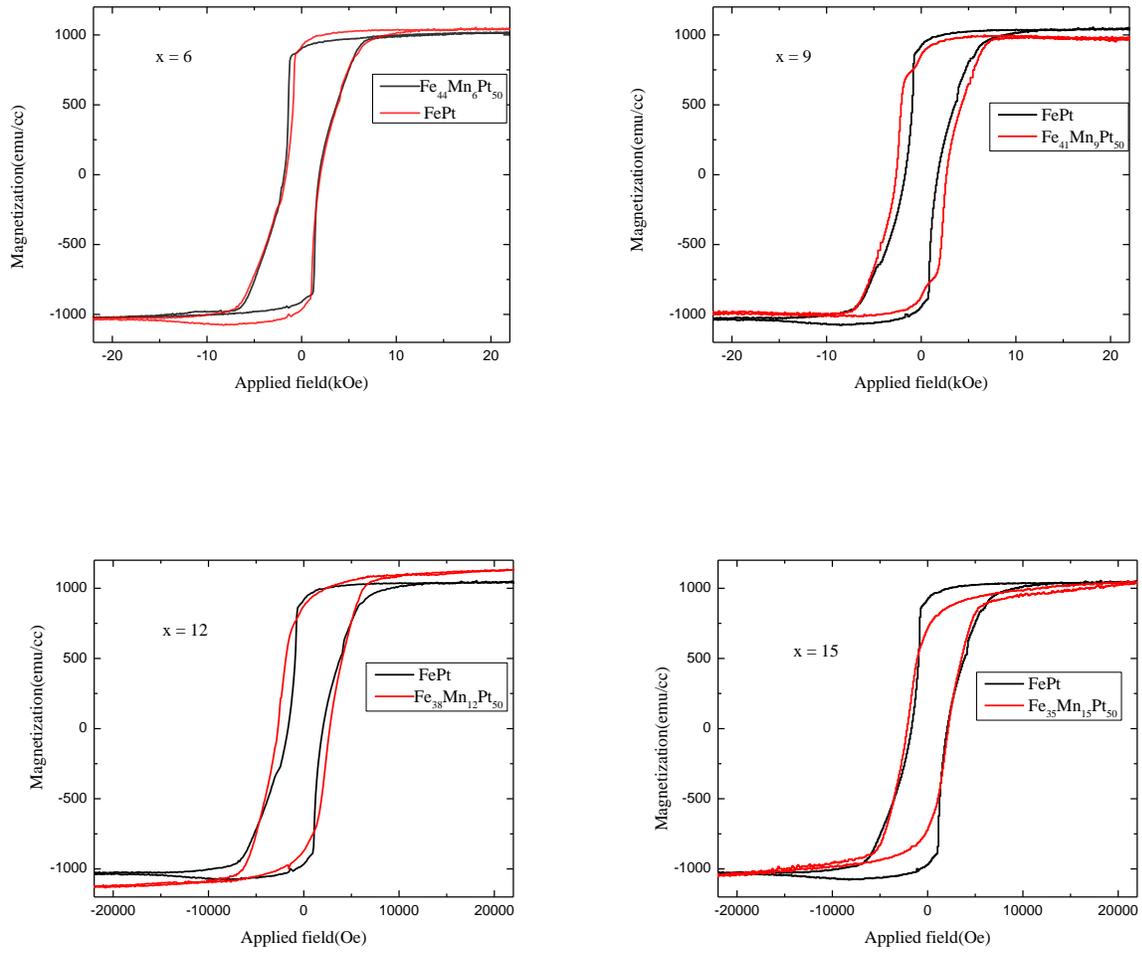

FIG. 4. Out of plane hysteresis loops for the $Fe_{50-x}Mn_xPt_{50}$ thin films

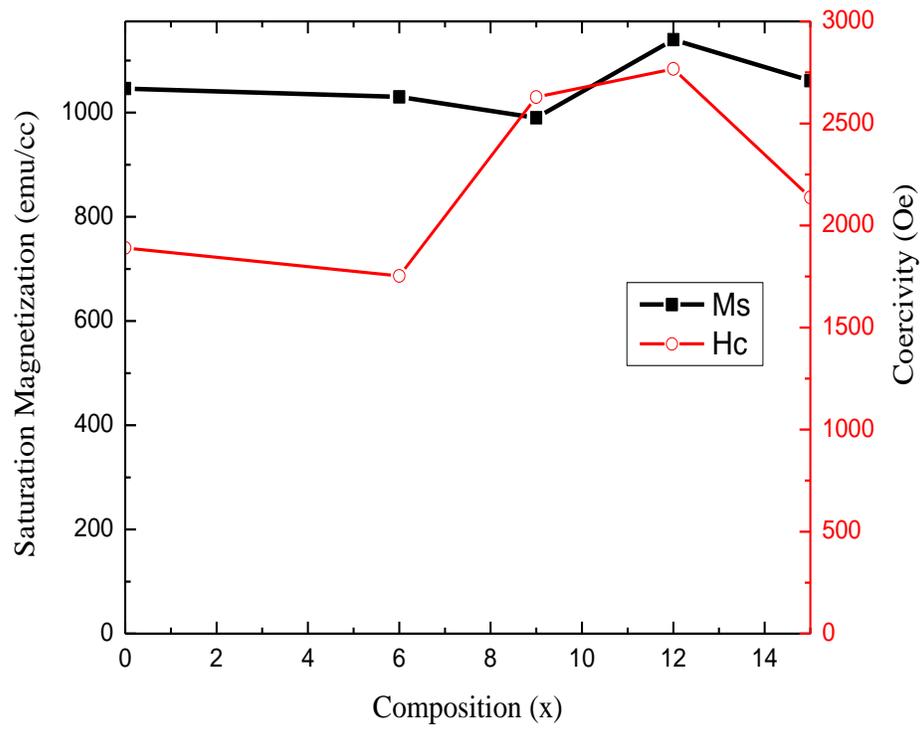

Figure 5. Saturation magnetization and out of plane coercivity as a function of composition for the $Fe_{50-x}Mn_xPt_{50}$ thin films on MgO (100)